\documentclass[aps,prl,twocolumn,amsmath,amssymb,superscriptaddress]{revtex4-2}

\usepackage[dvipsnames]{xcolor}
\usepackage{graphicx}
\usepackage{dcolumn}
\usepackage{bm}
\usepackage{url}
\usepackage[export]{adjustbox}
\usepackage{color}
\usepackage[normalem]{ulem} 

\usepackage[colorlinks=true,linkcolor=blue,citecolor=blue,urlcolor=blue]{hyperref}

\begin{document} 
   \title{Overdispersed radio source counts and excess radio dipole detection}
   \author{Lukas B\"ohme }\email{lboehme@physik.uni-bielefeld.de}
   \affiliation{Fakult\"at f\"ur Physik, Universit\"at Bielefeld,  Postfach 100131, 33501 Bielefeld, Germany}
   \author{Dominik J.~Schwarz} 
   \affiliation{Fakult\"at f\"ur Physik, Universit\"at Bielefeld,  Postfach 100131, 33501 Bielefeld, Germany}
   
    \author{Prabhakar Tiwari}
    \affiliation{Department of Physics, Guangdong Technion - Israel Institute of Technology, Shantou, Guangdong 515063, P.R. China}
    
    \author{Morteza Pashapour-Ahmadabadi} 
   \affiliation{Fakult\"at f\"ur Physik, Universit\"at Bielefeld,  Postfach 100131, 33501 Bielefeld, Germany}
    \author{Benedict Bahr-Kalus}
    \affiliation{INAF --  Istituto Nazionale di Astrofisica, Osservatorio Astrofisico di Torino, Via Osservatorio 20, 10025 Pino Torinese, Italy}
    \affiliation{Dipartimento di Fisica, Universit\`a degli Studi di Torino, Via P.\ Giuria 1, 10125 Torino, Italy}
    \affiliation{ INFN – Istituto Nazionale di Fisica Nucleare, Sezione di Torino, Via P.\ Giuria 1, 10125 Torino, Italy}
    
    \author{Maciej Bilicki}
    \affiliation{Center for Theoretical Physics, Polish Academy of Sciences, al. Lotników 32/46, 02-668 Warsaw, Poland}

    \author{Catherine L.~Hale}
    \affiliation{Astrophysics, Department of Physics, Denys Wilkinson Building, University of Oxford, Keble Road, Oxford, OX1 3RH, UK}
    
    \author{Caroline S.~Heneka}
     \affiliation{Institut f\"ur Theoretische Physik, Universit\"at Heidelberg, Philosophenweg 16, 69120 Heidelberg, Germany}
     
    \author{Thilo M.~Siewert}
      \affiliation{Fakult\"at f\"ur Physik, Universit\"at Bielefeld,  Postfach 100131, 33501 Bielefeld, Germany}
      \affiliation{Evangelisches Klinikum Bethel gGmbH, Kantensiek 11, 33615 Bielefeld, Germany}
        
   \date{\today}

\begin{abstract}
    The source count dipole from wide-area radio continuum surveys allows us to test the cosmological standard model. 
    Many radio sources have multiple components, which can cause an overdispersion of the source counts distribution.
    We account for this effect via a new Bayesian estimator, based on the negative binomial distribution.
    Combining the two best understood wide-area surveys, NVSS and RACS-low, and the deepest wide-area survey, LoTSS-DR2, we find that the source count dipole exceeds its expected value as the kinematic dipole amplitude from standard cosmology by a factor of $3.67 \pm 0.49$ --- a $5.4\sigma$ discrepancy.
\end{abstract}


\maketitle

\textit{Introduction}---The cosmological principle asserts that matter and radiation are isotropically and homogeneously distributed, assuming us to be typical observers. The dominant contribution to the photon number density in the Universe comes from the cosmic microwave background (CMB), a black body at $T_0 = 2.7$ K, which is observed to be highly isotropic. However, its primary deviation from isotropy is a temperature dipole with an amplitude of $T_1 = 3.4$ mK \cite{penzias1965,planckcollaboration2020a}, attributed to the motion of the solar system \citep{stewart1967, peebles1968}, with a speed of $v = (369.82\pm 0.11)\ \mathrm{km}/\mathrm{s}$ \cite{planckcollaboration2020a}.

The proper motion of an observer in an isotropic and homogeneous universe results in a dipole in the observed source counts \citep{ellis1984}. For a flux-limited radio continuum survey, this anisotropy arises from (i) a Doppler shift in frequency $\nu$, affecting the observed flux density $S_\nu$, modeled as $S_\nu \propto \nu^\alpha$, with spectral index $\alpha = \alpha(\nu)$, and (ii) aberration, causing source displacement towards the direction of motion. The kinematic source count dipole is given by \cite{ellis1984}
\begin{equation}
    \mathbf{d} = (2+x[1 - \alpha])\frac{\mathbf{v}}{c},\label{eq:d}
\end{equation}
where $x=x(\nu)$ is the slope of the cumulative source count at the flux density limit, $\mathrm{d} N/\mathrm{d} \Omega \left(>S_\nu\right) \propto S_\nu^{-x}$, $\mathbf{v}$ is the velocity, and $c$ the speed of light.

In addition, the observed source count dipole includes contributions from cosmic large-scale structures, known as the clustering dipole, and from shot noise due to the discrete nature of galaxy counts. Shot noise becomes less significant with a larger sample size. The clustering dipole depends on clustering strength, the growth factor, the redshift distribution of sources, and galaxy bias. In the flat $\Lambda$CDM model, the clustering dipole is typically smaller than the kinematic dipole if the sources are predominantly at redshifts greater than 0.1 \citep{tiwari2016,bengaly2019,cheng2023,oayda2023,dam2023,tiwari2024}.

To measure the dipole anisotropy, a large-sky survey is required. Radio continuum surveys have been key due to their broad sky coverage and detection of radio sources, which are typically at higher redshifts. Recently, infrared quasar surveys have also contributed to such measurements. These surveys, using catalogs of radio sources and quasars, have detected a significant dipole \cite{blakewall2002dipole,singal2011,gibelyou2012,rubart2013,siewert2021a,secrest2022a,mittal2024,wagenveld2023a}, which is inconsistent with the size of the CMB-predicted kinematic dipole. A notable discrepancy of up to 5.7 $\sigma$ \citep{dam2023} has been observed between velocities inferred from the CMB dipole and those from infrared sources though the dipole directions align at $(\alpha_{RA}, \delta) = (168^\circ, -7^\circ)$ in equatorial coordinates \cite{planckcollaboration2020a}. 

In contrast, few studies \cite{darling2022, wagenveld2024} find no discrepancy with the size of the CMB-predicted kinematic dipole. In \citep{darling2022}, the combination of two catalogues cancels their individual excess dipole amplitudes, as shown in \citep{secrest2022a}. The second study uses the sparsely sampled MeerKAT Absorption Line Survey Data Release 2 \cite{wagenveld2024}. The pointings exhibit systematic variation in source density of up to 5\%, as well as systematic trends with declination, both of which have been accounted for by an empirical fit.

Traditionally, studies have assumed that source counts follow a Poisson distribution, the standard model for shot noise, which works well for sparse data. However, high-resolution and high-sensitivity radio surveys challenge the independence of sources assumption due to overdispersion from multi-component sources. This overdispersion is significantly better modeled using the negative binomial distribution \citep{johnson2005}, providing a more accurate representation of the counts and associated noise.

In this letter, we present the first analysis of the radio source count dipole that accounts for overdispersion, including the deepest wide-area radio survey, the Low Frequency Array (LOFAR; \cite{LOFAR}) Two-metre Sky Survey Data Release 2 (LoTSS-DR2; \cite{LoTSS2}). Previous analyses of LoTSS-DR1 and LoTSS-DR2 excluded a Poissonian distribution with high statistical confidence, favoring a negative binomial distribution \cite{siewert2020,pashapour}. We extend this analysis to five other wide-area radio continuum surveys.

\textit{Data and theory}---The following six surveys, sorted by increasing central frequency, are examined in this letter: 

(1) The LoTSS-DR2 survey covers 5,635 deg$^2$ of the northern sky at 144 MHz, detecting nearly 4.4 million radio sources. It features a resolution of $6''$ and a median sensitivity of 83 $\mu$Jy beam$^{-1}$, with an astrometric accuracy of $0.2''$. The point source completeness is estimated at 95\% for 1.1 mJy. We use the inner masked regions as defined in \cite{hale2023a}.

(2) The TIFR GMRT Sky Survey, TGSS, alternative data release 1 \citep{intema2017}, conducted by the Giant Meterwave Radio Telescope (GMRT; \cite{GMRT}) covers the sky north of $\delta = -53^\circ$. 
It observes $\sim 0.62$ million sources at a central frequency of $147$\,MHz with a resolution of $25''$ for $\delta>19^\circ$, and $25'' \times 25''/\cos(\delta-19^\circ$) below. 
The median rms noise is $3.5$ mJy beam$^{-1}$, while $50\%$ point source completeness is expected to be reached at $25$ mJy. 
We mask $\delta < -49^\circ$ and $\delta > 80^\circ$ and remove three under- or overdense regions, as seen in Fig.~\ref{fig:data_overview} in the End Matter.

(3) The  Rapid ASKAP Continuum Survey, RACS \citep{mcconnell2020} is the first large sky survey with the Australian Square Kilometre Array Pathfinder (ASKAP; \cite{ASKAP,ASKAP2}). The first data release is RACS-low at  $887.5$\,MHz \cite{hale2021}.
The RACS-low survey covers the extragalactic sky ($|b|>5^\circ$) for $\delta \leq +30^\circ$ at a resolution of $25''$ and finds $\sim 2.1$ million radio sources. 
It is estimated to be 95\% point source complete at an integrated flux density of $\sim 3$ mJy. 

(4) The RACS-mid survey is the second data release of RACS at $1367.5$\,MHz \cite{duchesne2024} and covers the sky at $\delta \leq +49^\circ$. We use here the catalogue where images are convolved to a fixed $25''$ resolution which consists of $3$ million radio sources.
The estimated point source completeness is 95\% at $1.6$\,mJy. 
For both RACS releases (low and mid) we mask $\delta < -78^\circ$ and $\delta > +28^\circ$ as well as three and one small fields, respectively.

(5) The NRAO VLA Sky Survey, NVSS, \citep{NVSS} covers the sky for $\delta \geq -40^\circ$ at $1400$\,MHz and a resolution of $45''$. The completeness is given as $99\%$ at $3.4$ mJy. We mask $\delta < -39^\circ$ to prevent edge effects due to binning.

(6) The Very Large Array Sky Survey, VLASS, \citep{gordon2021, VLASS} observes the sky north of $\delta = -40^\circ$ in $S$ band ($2\text{\textendash}4$\,GHz) at a resolution of $2''\!.5$. The point source completeness of 99.8\% is reached at $3$\,mJy. A flux density correction of 10\% is applied to the cataloged flux densities from the first epoch \cite{gordon2021}. For masking we remove $\delta < -40^\circ$.

Additionally, for all surveys we mask the galactic plane at latitudes $|b| < 10^\circ$. Measurements are additionally affected by systematics, even above the completeness limit. To achieve a homogeneous sky coverage, flux density cuts up to an order of magnitude above the completeness limit are necessary \cite{wagenveld2023a,oayda2024a}.

The counts-in-cells distribution of continuum radio sources above a certain flux density is expected to follow a homogeneous Poisson point process \cite{neyman1954,peebles1980}. The probability of finding $N$ sources with an expected mean $\lambda$ is given by the Poisson distribution:
\begin{equation} 
    P_\mathrm{P}(N) = \frac{\lambda^N}{N!}\text{e}^{-\lambda}, 
\end{equation}
where the variance is $\sigma^2_\mathrm{P} = \lambda$.

However, the Poisson model cannot describe the observed overdispersion in source counts, as the high-resolution radio sky includes not only independent point sources but also multiple components from extended sources. Source-finding algorithms struggle to associate multi-component sources, and may count components as individual sources \cite{LoTSS2,hardcastle2023}. This results in overdispersion, where the observed variance of source counts exceeds the mean.
To address this, we adopt a compound Poisson (Cox) process \citep{cox1955}. A brief overview of the model is provided below; for details, see \citep{siewert2020,pashapour}. 

In any cell $i$, the number of independent radio objects is $O_i$, with each object having $C_{ji}$ components, where $j\in[1,\ldots,O_i]$. The total count of observed radio sources in cell $i$ is
\begin{equation} 
    N_i = \sum_{j=1}^{O_i} C_{ji}. 
\end{equation}
If $O_i$ follows a Poisson distribution with mean $\lambda$, and $C_{ji}$ follows a logarithmic distribution $\text{Log}(p)$ with $p\in(0,1)$ \cite{fisher1943}, this results in a negative binomial distribution. The probability of detecting $N_i$ radio sources in cell $i$ is \cite{pashapour}
\begin{equation} 
    P_\text{NB}(N_i) = \binom{N_i+r-1}{N_i} p^{N_i} (1-p)^r,\label{eq:pNB} 
\end{equation}
where $p$ encodes clustering, and $r$ relates to the Poisson mean $\lambda$ by $r=\lambda (1-p)/p$. The average number of components per object are given by the logarithmic distribution expectation:
\begin{equation} 
    \mu_\text{Log} = \frac{-1}{\ln(1-p)} \frac{p}{1-p}. 
\end{equation}
The mean and variance of the negative binomial distribution are
\begin{equation*} 
    \mu_{\text{NB}} = \frac{p\, r}{1-p}, \quad \sigma^2_{\text{NB}} = \frac{p\, r}{(1-p)^2}. 
\end{equation*}

So far, this model describes a uniform Universe with random fluctuations due to shot noise.
To include the radio source count dipole, we adopt the Bayesian approach that was proposed for the Poisson distribution \cite{wagenveld2023a}.
The source count dipole $\mathbf{d}$ is a variation of the expected source counts $\lambda_i$ for cell $i$, 
\begin{equation}
    \lambda_i =\lambda (1+d\cos\theta_i),
\end{equation} 
with $\theta_i$ the angle between the observed cell $i$ and the dipole direction. 
For the negative binomial distribution the parameter $r$ is directly linked to the direction-dependent quantity
\begin{equation}
    r_i = r(1+d\cos\theta_i). \label{eq:r_dipole}
\end{equation}

To estimate the source count dipole, we maximize the log-likelihood of the negative binomial dipole distribution 
\begin{align}
    \log \mathcal{L} =\sum_i &\bigl[ \log(\Gamma[N_i+ r_i]) 
     - \log(\Gamma[r_i]) \label{eq:estimatorNB} \\
     &+ r_i
     \log(1-p) + C_i \bigr] \nonumber ,
\end{align}
where $C_i= N_i \log p - \log(\Gamma[N_i+1])$, and $\Gamma[\cdot]$ denotes the gamma function. 
We use a Markov Chain Monte Carlo  method to estimate  $r$, $d$ and the direction $\theta(\alpha_{RA}, \delta)$, while p is inferred from the empirical mean and variance via $p=1-\widehat\mu/\widehat{\sigma^2}$.

Table~\ref{tab:comp} lists the number of cells, reduced $\chi_r^2$ values of the best-fit Poisson and negative binomial distributions, and the inferred mean number of components $\mu_\text{Log}$ for each radio continuum surveys. Figure~\ref{fig:data_overview} in the End Matter displays the corresponding masked survey maps, counts-in-cells histograms, and comparisons between the two best-fit distributions. The negative binomial distribution provides a significant better fit to the overdispersed data across all surveys.

\begin{table}
\caption{Comparison of counts-in-cells between best-fit Poisson and negative binomial distributions as seen in Fig.~\ref{fig:data_overview} in the End Matter. Cells cover a sky area of $\sim3.4~\text{deg}^2$.\label{tab:comp}}
\bgroup\renewcommand{\arraystretch}{1.2}
\begin{tabular}{lccccc}\hline\hline
Survey & \# Cells & $S_\mathrm{min} [\text{mJy}]$ & $\chi^2_\mathrm{r,P}$ & $\chi^2_\mathrm{r,NB}$ & $\mu_\text{Log}$\\ \hline
 LoTSS-DR2 & 1267 &  $5$   & $2.25$ & $0.76$ & $1.70 \pm 1.28$ \\
 TGSS      & 8395 & $100$  & $23.79$ & $0.87$ & $1.22 \pm 0.55$\\
 RACS-low  & 7347 & $20$   & $4.51$ & $1.18$ & $1.10 \pm 0.33$ \\ 
 RACS-mid  & 7365 & $20$   & $2.80$ & $0.70$ & $1.07 \pm 0.27$ \\
 NVSS      & 8364 & $20$   & $4.14$ & $1.09$ & $1.09 \pm 0.32$ \\ 
 VLASS     & 8369 & $10$   & $40.9$ & $0.90$ & $1.32 \pm 0.71$ \\ \hline\hline
\end{tabular}
\egroup
\end{table}

\textit{Results}---We use the {\sc HEALPix}\footnote{http://healpix.sourceforge.net}
binning scheme with $N_\mathrm{side} = 32$, yielding $12288$ equal-area sky cells. Log-likelihoods are maximized using \textsc{Bilby} \citep{bilby1,bilby2} with the \texttt{emcee} sampler \citep{emcee}. Estimator accuracy was verified through numerical simulations with $3\times 10^5$ sources over the extragalactic sky ($-20^\circ < \delta < 90^\circ$), matching the source count and sky coverage of the surveys used.

To find the expected dipole amplitude $d_{\text{exp}}$ from the inferred CMB dipole velocity, we measure $x$ and $\alpha$ for each survey and flux density cut. $x$ is measured over a narrow flux range ($\mathcal{O}(1~\text{mJy})$) around the flux density cut to ensure a good fit. For $\alpha$, a simple positional cross-matching method is applied between surveys, with a search radius equal to half the resolution of NVSS, $45''/2$;  a flux density cut is applied to the survey for which $\alpha$ is calculated.
The expected dipole amplitude is calculated using Eq.~\eqref{eq:d} and the results are summarized in Table~\ref{tab:x}. More details can be found in Table~\ref{tab:alpha} in the End Matter.

\begin{table}
\caption{Expected dipole amplitude $d_{\text{exp}}$ using \eqref{eq:d}, the flux density cut, the measured spectral index $\alpha$ and the corresponding differential source count slope $x$. Errors for $x$ are smaller than $10^{-3}$ and therefore not listed, but used in the error calculation of  $d_{\text{exp}}$. \label{tab:x}}
\bgroup\renewcommand{\arraystretch}{1.2}
\begin{tabular}{lccccc}\hline\hline
Survey & $S_\mathrm{min}$ & $x$ & $\alpha$ & $d_{\text{exp}}$\\ 
& (mJy) & & & ($\times 10^{-2}$)\\ \hline
 LoTSS-DR2  & $5$   & $0.74$ & $-0.73\pm0.23$ & $0.405 \pm 0.021$\\
 TGSS       & $100$ & $0.79$ & $-0.76\pm0.18$ & $0.418 \pm 0.018$ \\ 
 RACS-low   & $20$  & $0.85$ & $-0.98\pm0.24$ & $0.454 \pm 0.025$\\ 
 RACS-mid   & $20$  & $0.88$ & $-0.81\pm0.30$ & $0.443 \pm 0.025$\\ 
 NVSS       & $20$  & $0.88$ & $-0.73\pm0.23$ & $0.435 \pm 0.025$ \\
 VLASS      & $10$  & $0.99$ & $-0.71$\footnote{taken from \citep{gordon2021}, which used the Faint Images of the Radio Sky at Twenty-Centimeters (FIRST; \cite{FIRST}) to calculate $\alpha$.}        & $0.456 \pm 0.037$\\ \hline\hline
\end{tabular}
\egroup
\end{table}
The novel estimator {introduced} in Eq.~\eqref{eq:estimatorNB} is applied to each survey at different flux density cut.
A uniform prior is used for both dipole direction and amplitude $d$, labeled as `Free' in Table~\ref{tab:comp_poisson_results}. Restricting the direction to the CMB dipole direction is labeled as `CMB'.
The `Free' results fall into two categories: (i) `Problematic': LoTSS-DR2, RACS-mid and VLASS show unreliable $\delta$ retrievals with TGSS yielding an unusually high amplitude; and (ii) Robust: RACS-low and NVSS provide stable and robust measurements, as listed in Table~\ref{tab:comp_poisson_results} under `Free'.

Figure~\ref{fig:dec_corners} in the End Matter illustrates $\delta$ behavior across flux density cuts. LoTSS-DR2 and VLASS tend toward the celestial poles, while RACS-mid shows strong flux-dependent variations in $\delta$. For LoTSS-DR2, this is expected due to limited sky coverage and is confirmed via simulations with the survey mask. Using random mocks \cite{hale2023a} and injecting a dipole as in Table 3, we find a pronounced variance of $\delta$.
VLASS consistently favors $\delta \geq 60^\circ$, regardless of flux cut, indicating systematic issues, likely from declination-dependent effects. As VLASS is based on \textit{Quick Look} images, it lacks the precision of fully calibrated surveys. Dipoles near the poles can result from declination-dependent systematics in telescope sensitivity or calibration. For RACS-low and NVSS, we confirm previous results \citep{wagenveld2023a} showing that the excess dipole is about $3.5$ times the expected CMB dipole amplitude. 

We confirm the increased dipole in TGSS, which is roughly ten times larger than the CMB expectation \citep{siewert2021a}. This excess likely stems, at least partly, from large-scale systematic deviations in flux density calibration \citep{tiwari2019}. However, an improved catalogue \cite{hurley-walker2017a} does not yield a lower amplitude.

Next, we constrain the estimator's direction to the CMB dipole to measure the amplitude projected along that axis, applying robust flux density cuts for each survey, well above the 95\% completeness limits and in line with prior works. For LoTSS-DR2 and RACS-mid, the constrained estimator yields results consistent with the reported dipole excess of $\approx 3 \times d_\textrm{exp}$.
However, these values should be considered lower bounds, since the true dipole may deviate by several tens of degrees from the CMB dipole direction. This is evident in VLASS, where the constrained amplitude matches expectations, but the actual orientation differ by $60^\circ - 80^\circ$ (see Fig.~\ref{fig:dec_corners} in the End Matter.)

All measurements are repeated with the Poisson estimator \cite{wagenveld2023a} for which the results are also provided in Table~\ref{tab:comp_poisson_results}.
Differences between the Poisson and negative binomial estimators mainly appear in the error bars, which vary significantly depending on survey properties such as resolution and sensitivity.
For instance, in LoTSS-DR2, the standard deviation under the negative binomial model is $0.44\times10^{-2}$, compared to $0.27\times10^{-2}$ for the Poisson case, an increase of around 60\%.

As the final step of our analysis, we combine the two wide-area surveys, RACS-low and NVSS, which are the only ones yielding stable results with the unconstrained estimator, both alone and together with the deeper LoTSS-DR2 survey (without fixing the dipole direction).

\begin{figure}
    \includegraphics[width=.999\linewidth]{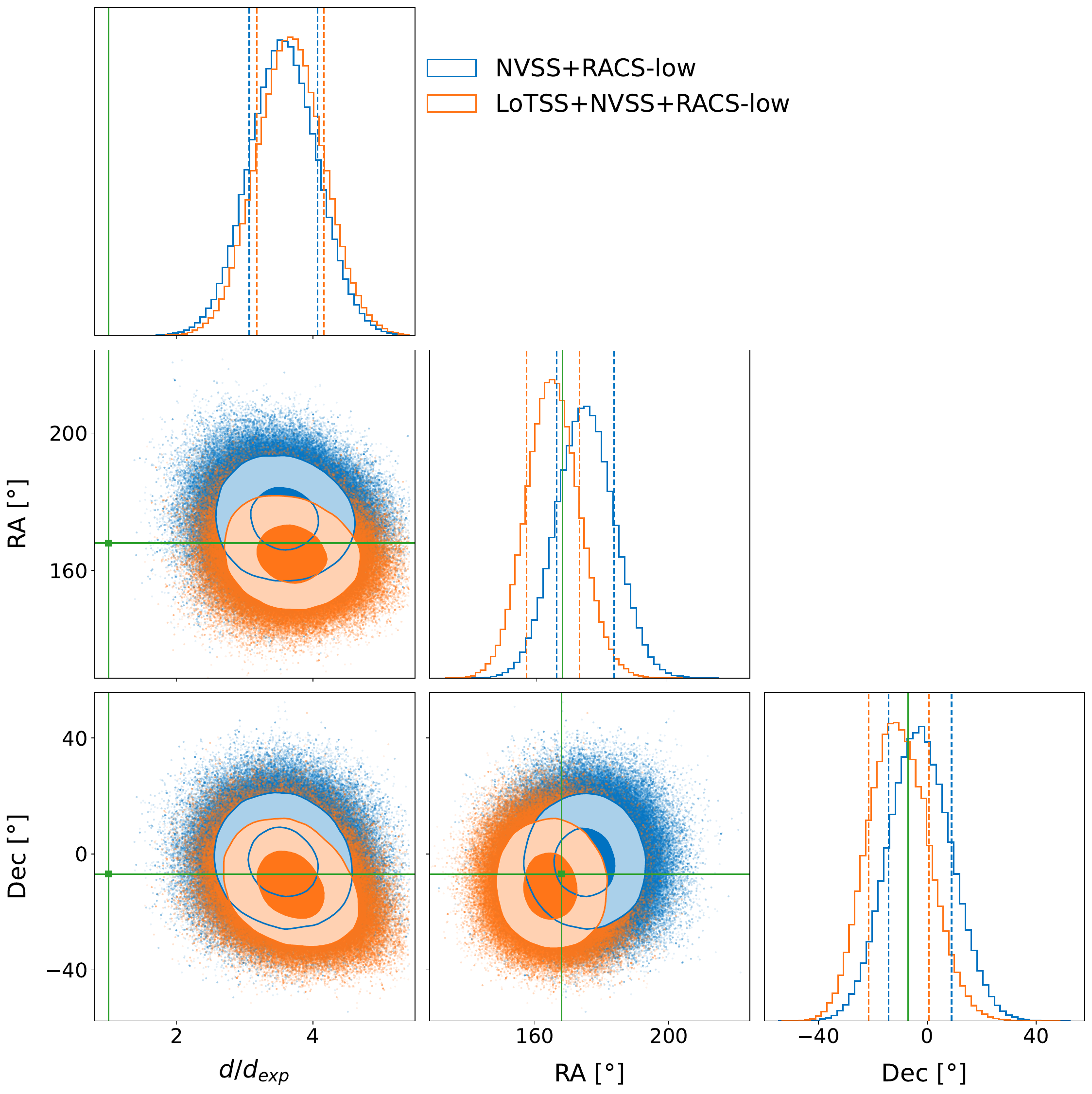}
    \caption{Corner plot for the combined dipole estimate from LoTSS-DR2, NVSS, and RACS-low. The amplitude is presented in multiples of the expected dipole amplitude, with the green lines and dot representing the expected values based on the CMB dipole.}
    \label{fig:Dipole_combined}
\end{figure}

As shown in Fig.~\ref{fig:Dipole_combined} and detailed in Table~\ref{tab:comp_poisson_results}, the combined measurement from all three surveys yields a direction closely aligned with the CMB dipole ($\alpha_{RA}, \delta$) = ($165^\circ\pm8^\circ, -11^\circ\pm11^\circ$), with an angular separation $\Delta \theta = (5\pm10)^\circ$, and amplitude $(3.67 \pm 0.49)\times d_\textrm{exp}$. Adding LoTSS-DR2 increases the discrepancy in amplitude from $4.8\sigma$ (using the Poisson estimator on RACS-low and NVSS \cite{wagenveld2023a}) to over $5.4\sigma$ with our negative binomial estimator. Applying the negative binomial estimator to RACS-low and NVSS alone (with the same parameters as \cite{wagenveld2023a}) reduces the tension to $4.5\sigma$, with $\Delta \theta = (8\pm10)^\circ$, as expected due to the larger error bars.

\begin{table*}[]
    \centering
    \caption{Measurement results of amplitude $d$ and direction ($\alpha_{RA}$,$\delta$) of the radio source count dipole and comparison between Poisson (P) and negative binomial distribution (NB). $S_\text{min}$ is the applied minimum flux density cut at the survey's frequency, while this value scaled to $144$ MHz is given as $S_{144}$ and calculated with a general spectral index of $-0.75$. The direction `Free' refers to an unconstrained estimator, with the preferred direction given in the columns ($\alpha_{RA}$,$\delta$). `CMB' refers to the constrained estimator, where the direction is fixed to the CMB dipole direction. Values without error bars were fixed during parameter estimation.}
    \bgroup\renewcommand{\arraystretch}{1.5}%
    \begin{tabular}{lcccccrrrrrr}\hline\hline
    Survey   & $\nu$ & $S_\text{min}$ & $S_{144}$ & N & Direction  & $d$ (P)  & $d$ (NB)      & $\alpha_{RA}$ (P)   & $\alpha_{RA}$ (NB)   & $\delta$ (P)   & $\delta$ (NB) \\ 
     & (MHz) & (mJy) & (mJy) & & & $(\times 10^{-2})$ &  $(\times 10^{-2})$ & (deg) & (deg) & (deg) &  (deg) \\ \hline 
    LoTSS-DR2 & 144 & 5 & 5 & 376658  & CMB         & $1.36^{+0.24}_{-0.30}$  & $1.39^{+0.45}_{-0.42}$  & 167.94     & 167.94    & -6.94       & -6.94    \\[5pt]
    TGSS      & 147 & 100 & 102 & 234619 & Free       & $5.33^{+0.34}_{-0.36}$  &  $5.33^{+0.40}_{-0.42}$&  $140.7^{+3.5}_{-3.7}$  &   $140.7^{+4.4}_{-4.4}$   &   $1.5^{+4.6}_{-4.4}$   & $1.6^{+5.6}_{-5.2}$    \\ 
              &     & 100 &   & & CMB        & $4.51^{+0.30}_{-0.34}$ & $4.51^{+0.36}_{-0.40}$ & 167.94     & 167.94    & -6.94       & -6.94   \\[5pt]
    RACS-low  & 888 & 20 & 78 & 330540 & Free        & $1.78^{+0.26}_{-0.26}$  & $1.80^{+0.28}_{-0.30}$ & $191.1^{+ 9.7}_{ -9.8}$ & $191.1^{+ 10.8}_{-10.8}$ & $  5.8^{+ 13.8}_{-12.5}$  & $  6.0^{+ 14.2}_{-14.1}$ \\ 
              &     & 20 &   & & CMB         & $1.70^{+0.24}_{-0.24}$  & $1.73^{+0.27}_{-0.30}$  & 167.94     & 167.94    & -6.94       & -6.94    \\[5pt]
    RACS-mid  & 1367 & 20 & 108 & 237069 & CMB          & $1.15^{+0.30}_{-0.30}$ & $1.15^{+0.33}_{-0.33}$   & 167.94     & 167.94    & -6.94       & -6.94     \\[5pt]
    NVSS      & 1400 & 20 & 110 & 263125 & Free        & $1.54^{+0.30}_{-0.34}$  & $1.54^{+0.36}_{-0.36}$  & $148.4^{+ 12.6}_{-12.5}$  & $148.1^{+ 13.9}_{-13.1}$  & $-10.3^{+ 14.7}_{-14.4}$ & $-9.5^{+ 16.0}_{-15.6}$ \\ 
              &      & 20 &  & & CMB         & $1.36^{+0.27}_{-0.30}$  & $1.36^{+0.30}_{-0.30}$ & 167.94     & 167.94    & -6.94       & -6.94    \\[5pt]
    VLASS     & 3000 & 10 & 97.5 & 275516 & CMB          & $0.55^{+0.27}_{-0.30}$ & $0.55^{+0.36}_{-0.30}$    & 167.94     & 167.94    & -6.94       & -6.94   \\[10pt] 
    \multicolumn{5}{l}{Survey combination} & Direction  & $d/d_{exp}$ (P)  & $d/d_{exp}$ (NB)      & $\alpha_{RA}$ (P)   & $\alpha_{RA}$ (NB)   & $\delta$ (P)   & $\delta$ (NB) \\ \hline
    \multicolumn{5}{l}{RACS-low + NVSS} & Free & $3.55^{+0.46}_{-0.46}$ & $3.57^{+0.50}_{-0.50}$ & $175.4^{+ 8.3}_{-8.1}$ & $175.0^{+ 9.0}_{-8.9}$ & $-3.0^{+ 10.5}_{-10.5}$ & $-2.8^{+ 11.8}_{-11.3}$ \\ 
    \multicolumn{5}{l}{LoTSS-DR2 + RACS-low + NVSS} & Free & $3.92^{+0.46}_{-0.45}$ & $3.67^{+0.49}_{-0.49}$ & $156.1^{+ 6.8}_{-6.7}$ & $164.9^{+ 8.3}_{-8.0}$ & $-20.1^{+ 9.6}_{-8.7}$ & $-10.8^{+ 11.4}_{-10.8}$ \\  \hline \hline
    \end{tabular}
    \egroup%
    \label{tab:comp_poisson_results}
\end{table*}

\textit{Conclusion}---In this work, we develop a new radio dipole estimator based on the negative binomial distribution, motivated by the observed source count distribution. It is applied to six different radio surveys spanning $120\,$MHz to $4$\,GHz, covering most of the northern or southern sky. 

Combining two wide-area surveys (RACS-low, NVSS) with the deeper LoTSS-DR2, our estimator robustly constrains the dipole amplitude to $(3.67 \pm 0.49)\times d_\mathrm{exp}$, revealing a $5.4\sigma$ excess over the CMB-inferred kinematic dipole. The direction remains consistent with the CMB dipole within $1\sigma$. This marks the first significant detection of the radio dipole excess using radio surveys alone.

We find that the multi-component nature of radio sources, captured by our estimator, is a key factor in measuring the cosmic radio dipole and can increase the amplitude uncertainty by up to 60\%.

Possible explanations for the excess radio source count dipole include greater-than-expected contamination from local sources, potentially producing a clustering dipole  not predicted by $\Lambda$CDM \cite{tiwari2016,bengaly2019}. A local bulk flow extending beyond $\Lambda$CDM expectations may also contribute \cite{dipolerepeller,watkins2023a,Courtois2023}.

Another possible explanation is systematics, extensively studied across many surveys (see Appendix B in the End Matter for references and details). Achieving precise flux density calibration over wide fields is challenging and may introduce bias.
An alternative strategy is sparse sky sampling via independent pointings, which yields a dipole consistent with the CMB expectation \cite{wagenveld2024}; however, unlike the active galactic nuclei-dominated surveys in this work, that sample is dominated by star-forming galaxies.

Survey geometry, whether sparse or wide, can itself bias dipole measurements and must be tested via simulations. Galactic synchrotron emission may affect sensitivity if not properly addressed during calibration. However, it is unlikely that these systematics would be consistent across all radio frequencies or mimic results seen in infrared quasars studies \cite{secrest2021}.

A genuine discrepancy between the dipole amplitudes measured in the CMB and large-scale structure frames would have profound cosmological implications. Depending on whether the cause is local or large-scale, it could challenge the cosmological principle itself.

Upcoming large-area sky surveys such as LoTSS-DR3, LoLSS-DR2 (northern sky at $54$\,MHz), RACS-high \cite{duchesne2025}, RACS-low-DR2, EMU \citep{EMU,EMU2}, and eventually the SKA surveys, along with wide-field spectroscopic follow-ups like the WEAVE-LOFAR project \citep{smith2016}, will significantly improve our understanding of the origin of the radio dipole.\\

\begin{acknowledgments}
\textit{Acknowledgments}---We acknowledge discussions with Nathan Secrest, Sebastian von Hausegger, and Jonah Wagenveld who helped us to sharpen and develop our arguments. 
LB acknowledges support by the Studienstiftung des deutschen Volkes. 
MPA acknowledges support from the Bundesministerium für Bildung und Forschung (BMBF) ErUM-IFT 05D23PB1.
BB-K acknowledges support from INAF for the project `Paving the way to radio cosmology in the SKA Observatory era: synergies between SKA pathfinders/precursors and the new generation of optical/near-infrared cosmological surveys' (CUP C54I19001050001).
MB is supported by the Polish National Science Center through grants no. 2020/38/E/ST9/00395 and 2020/39/B/ST9/03494.
CLH acknowledges support from the Oxford Hintze Centre for Astrophysical Surveys which is funded through generous support from the Hintze Family Charitable Foundation.
CSH’s work is funded by the Volkswagen Foundation.
CSH acknowledges additional support from the
Deutsche Forschungsgemeinschaft (DFG, German Research Foundation) under Germany’s Excellence Strategy EXC 2181/1 - 390900948 (the Heidelberg STRUCTURES Excellence Cluster).
LOFAR is the Low Frequency Array designed and constructed by ASTRON. It has observing, data processing, and data storage facilities in several countries, which are owned by various parties (each with their own funding sources), and which are collectively operated by the ILT foundation under a joint scientific policy. The ILT resources have benefited from the following recent major funding sources: CNRS-INSU, Observatoire de Paris and Université d’Orléans, France; BMBF, MIWF-NRW, MPG, Germany; Science Foundation Ireland (SFI), Department of Business, Enterprise and Innovation (DBEI), Ireland; NWO, The Netherlands; The Science and Technology Facilities Council, UK; Ministry of Science and Higher Education, Poland; The Istituto Nazionale di Astrofisica (INAF), Italy.
This scientific work uses data obtained from Inyarrimanha Ilgari Bundara/the Murchison Radio-astronomy Observatory. We acknowledge the Wajarri Yamaji People as the Traditional Owners and native title holders of the Observatory site. CSIRO’s ASKAP radio telescope is part of the Australia Telescope National Facility (https://ror.org/05qajvd42). Operation of ASKAP is funded by the Australian Government with support from the National Collaborative Research Infrastructure Strategy. ASKAP uses the resources of the Pawsey Supercomputing Research Centre. Establishment of ASKAP, Inyarrimanha Ilgari Bundara, the CSIRO Murchison Radio-astronomy Observatory and the Pawsey Supercomputing Research Centre are initiatives of the Australian Government, with support from the Government of Western Australia and the Science and Industry Endowment Fund.
The GMRT is run by the National Centre for Radio Astrophysics of the Tata Institute of Fundamental Research. 
The VLA is run by the National Radio Astronomy Observatory, a facility of the National Science Foundation operated under cooperative agreement by Associated Universities, Inc.
We used a range of python software packages during this work and the production of this manuscript, including Astropy \citep{Astropy1,Astropy2,Astropy3}, matplotlib \citep{Matplotlib}, NumPy \citep{Numpy}, SciPy \citep{Scipy}, healpy \citep{Healpy}, and {\sc HEALPix} \citep{Healpix}.
\end{acknowledgments}


%

\begin{figure}
    \includegraphics[width=.77\linewidth]{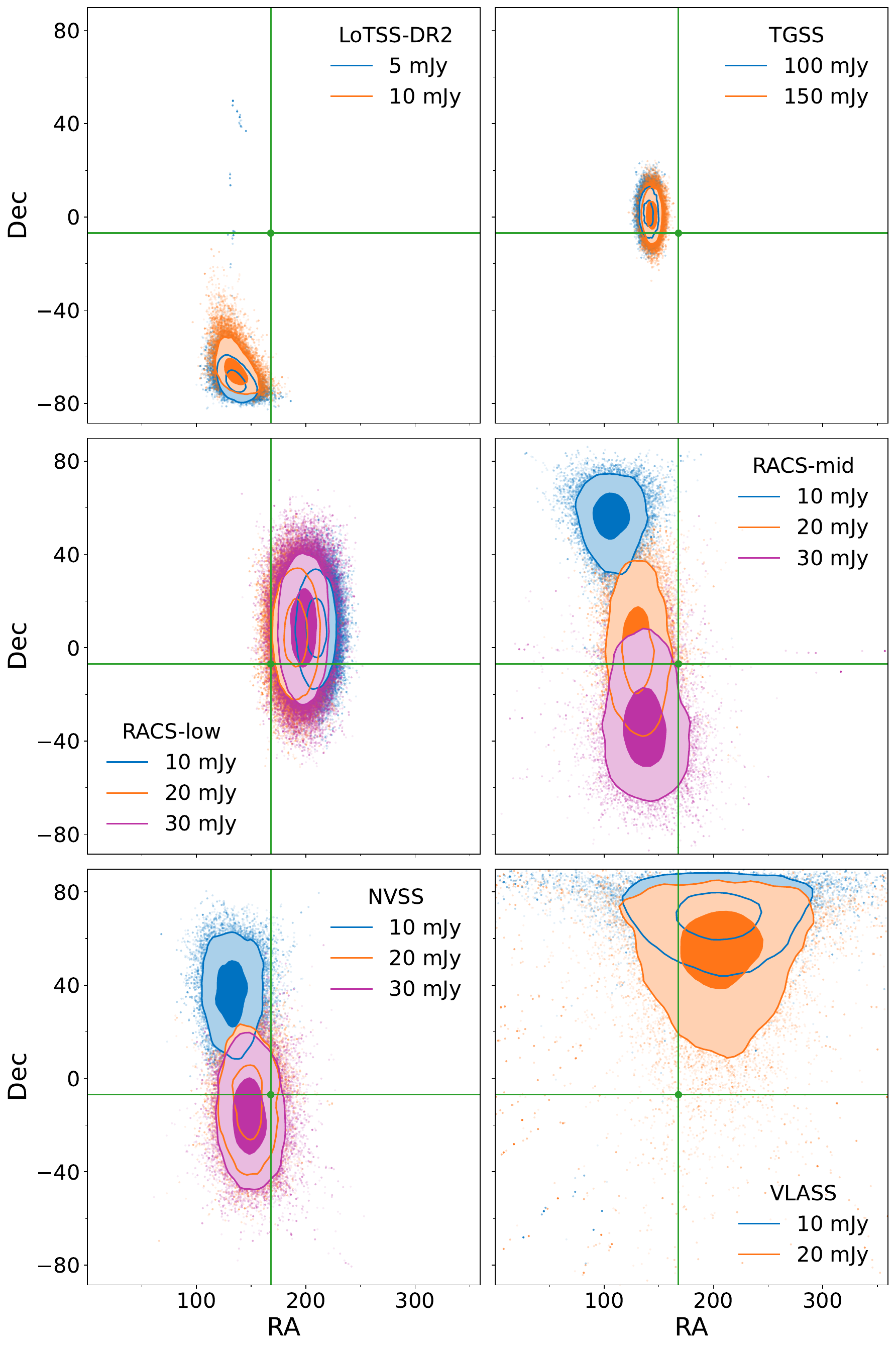}
    \caption{Declination and right ascension of source count dipole for all six radio surveys for different flux density cuts using the unconstrained estimator. The green lines and dot represent the expected value based on the CMB dipole.}
    \label{fig:dec_corners}
\end{figure}

\clearpage 
\appendix
\begin{widetext}
\begin{figure*}
        \includegraphics[width=.32\linewidth,valign=t]{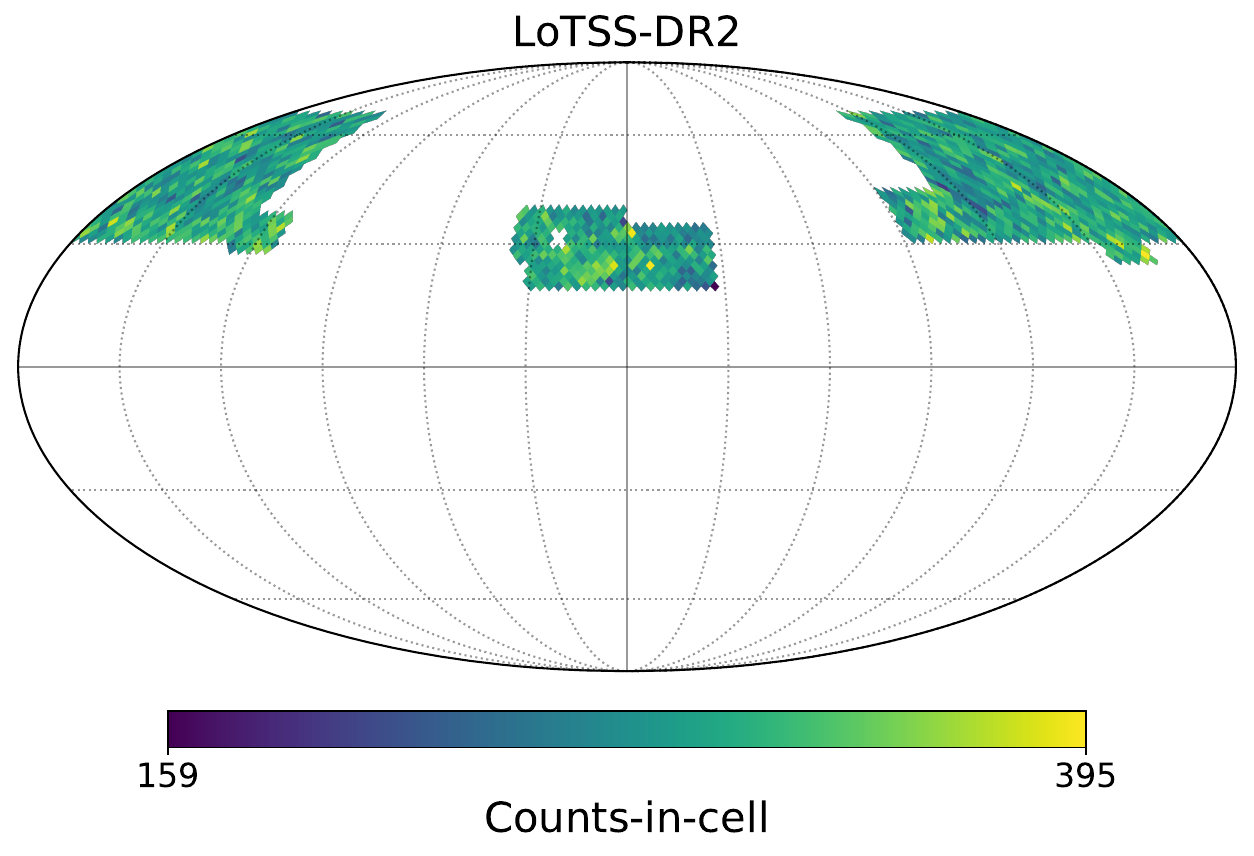}
        \includegraphics[width=.32\linewidth,valign=t]{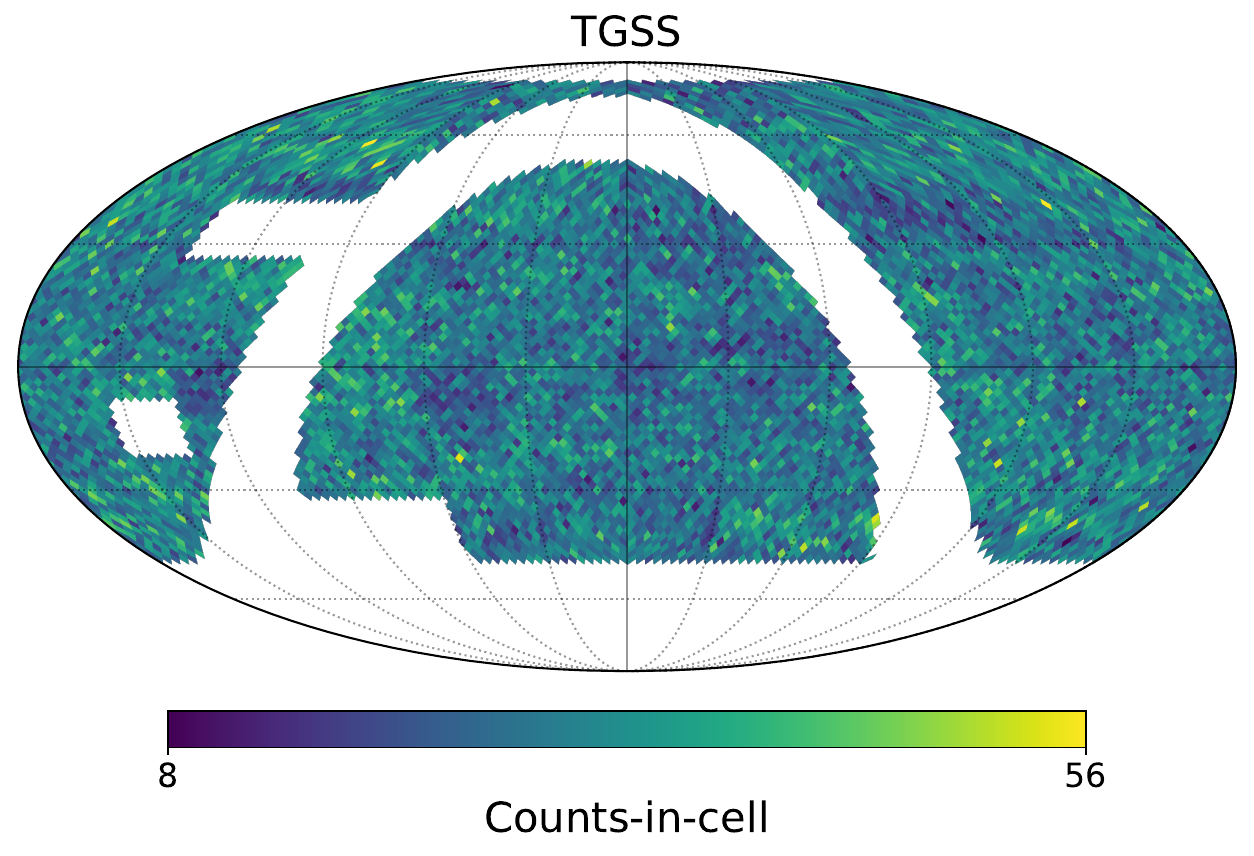}
        \includegraphics[width=.32\linewidth,valign=t]{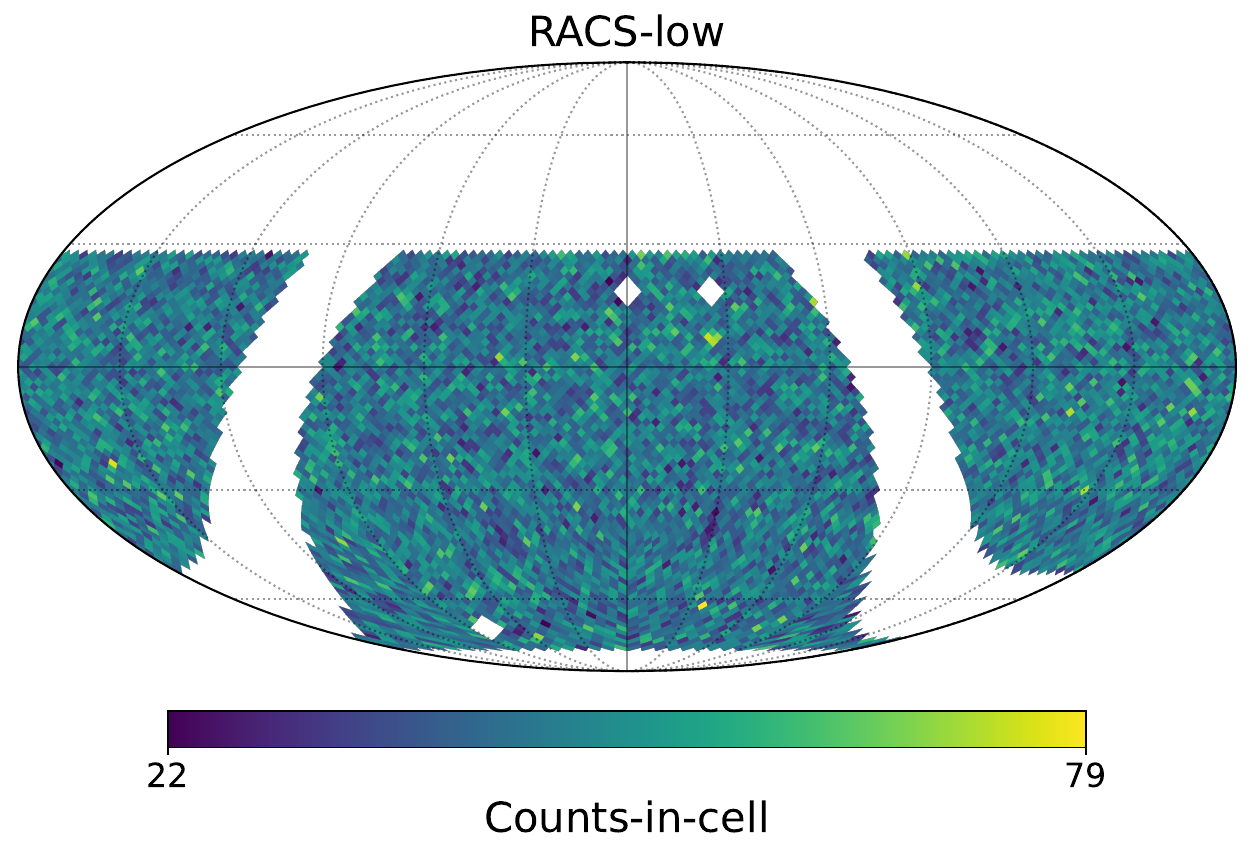}\\
        \includegraphics[width=.32\linewidth]{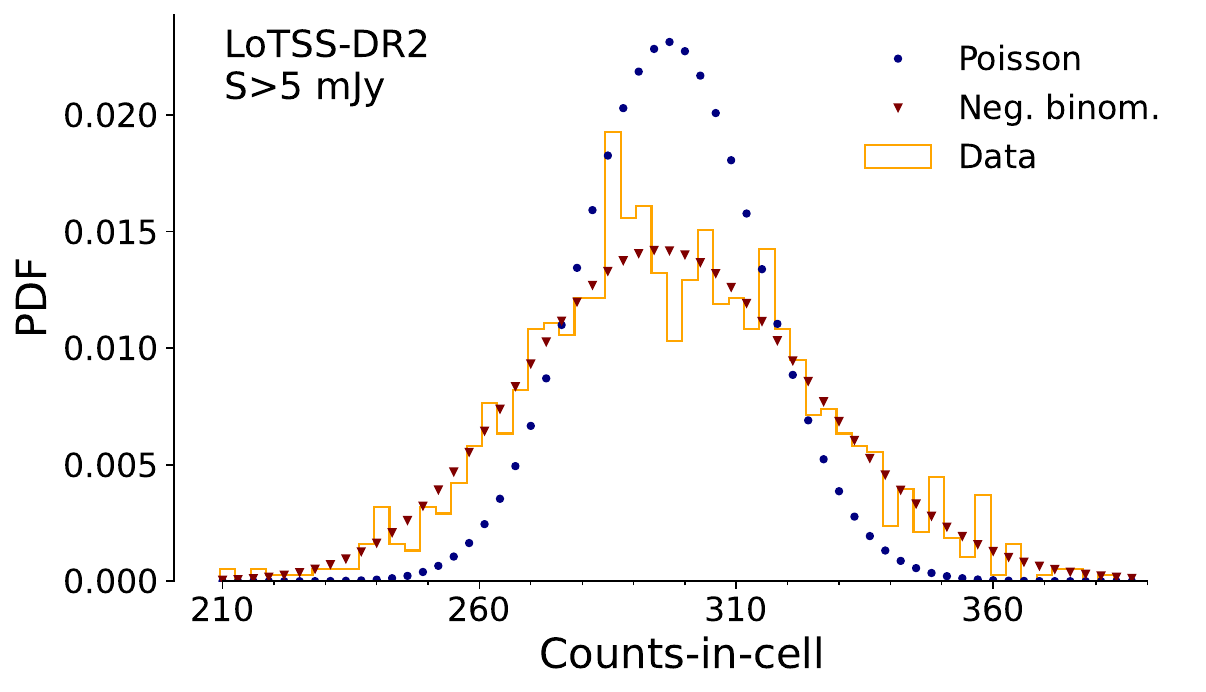}
        \includegraphics[width=.32\linewidth]{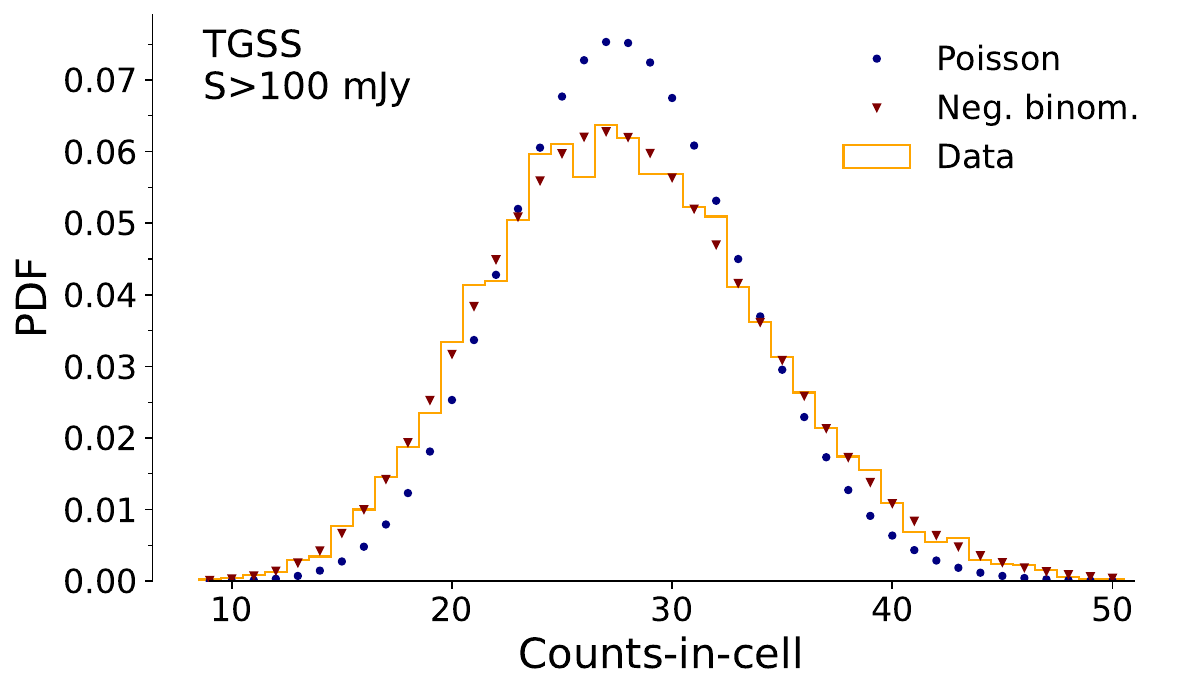}
        \includegraphics[width=.32\linewidth]{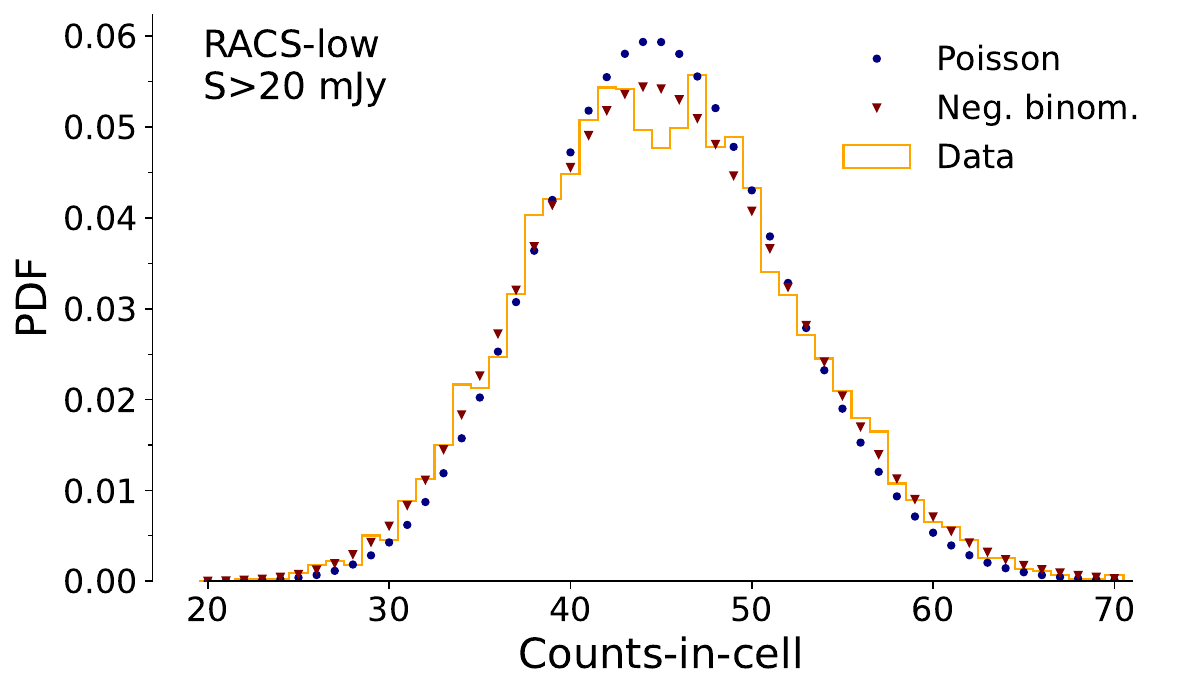} \\
        \includegraphics[width=.32\linewidth,valign=t]{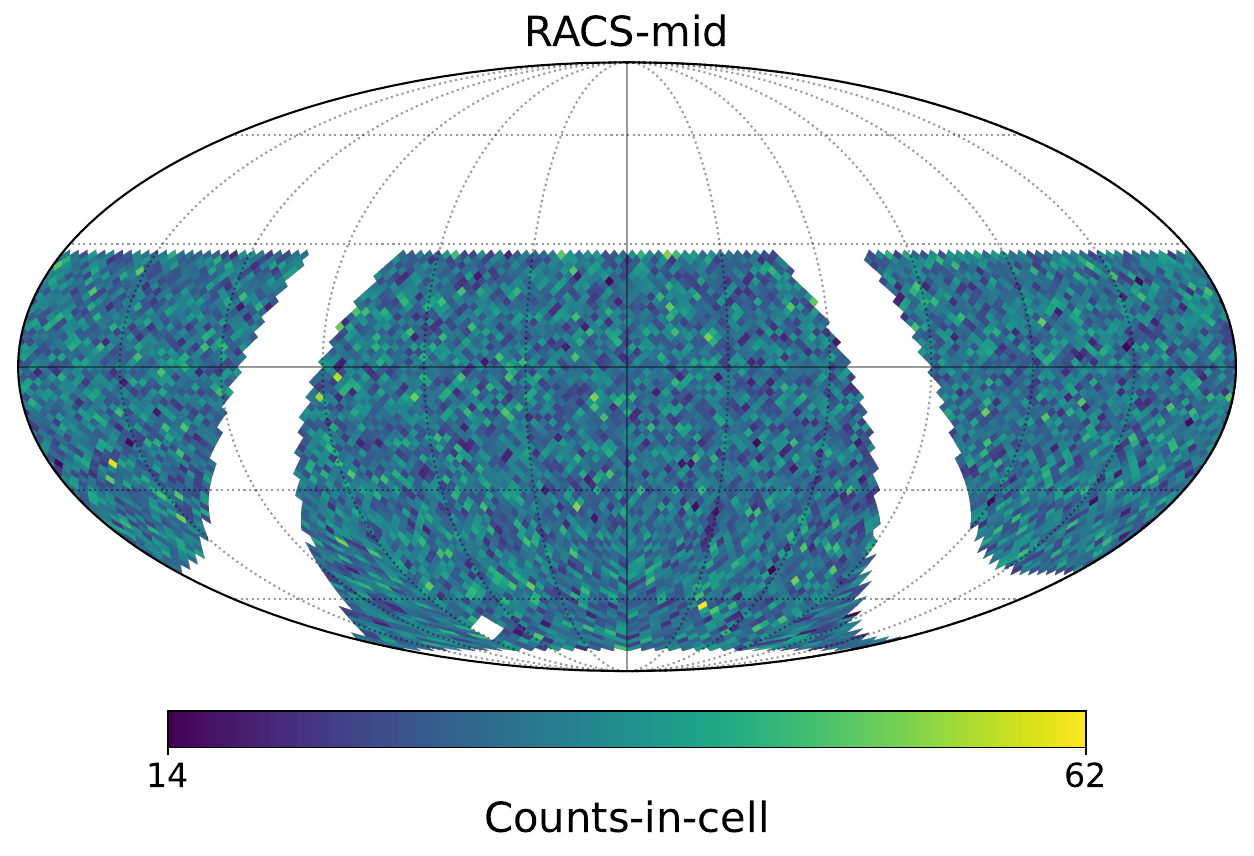}
        \includegraphics[width=.32\linewidth,valign=t]{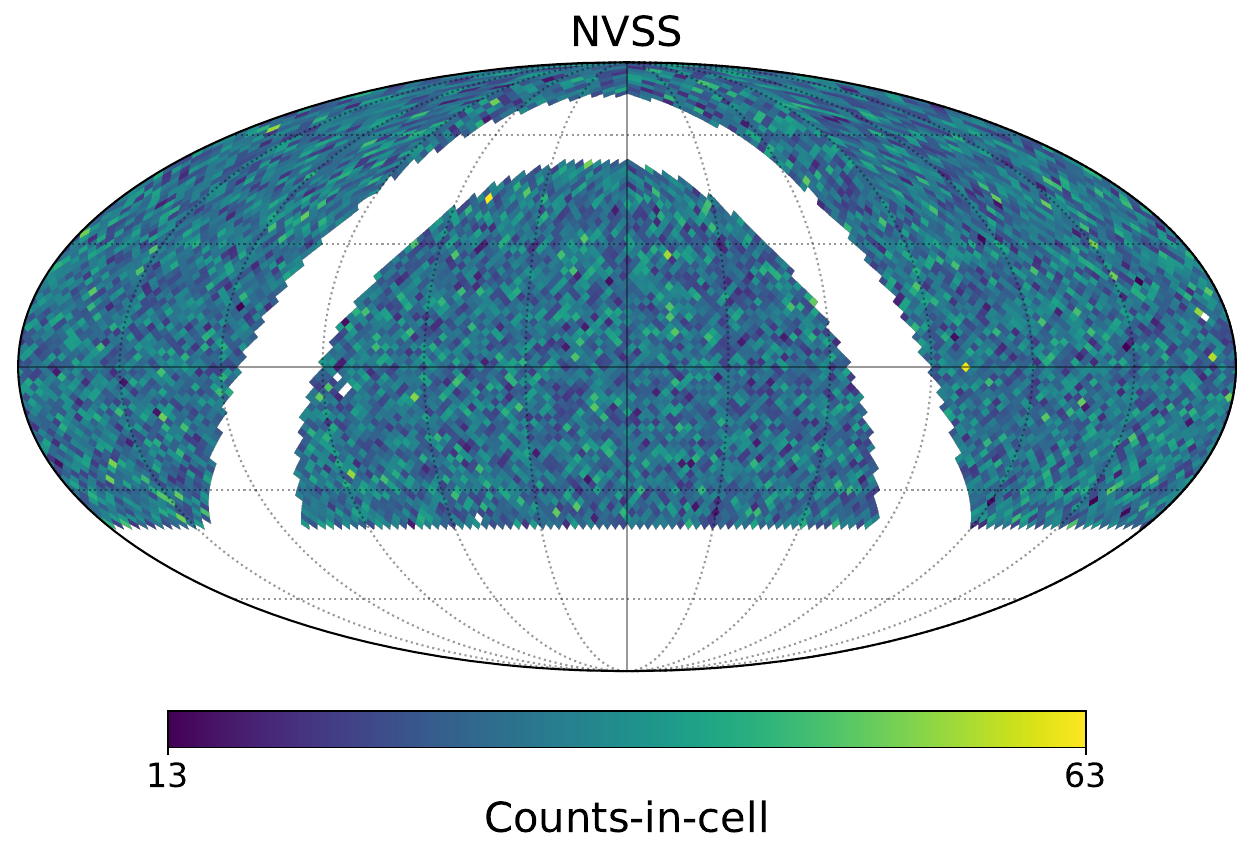}
        \includegraphics[width=.32\linewidth,valign=t]{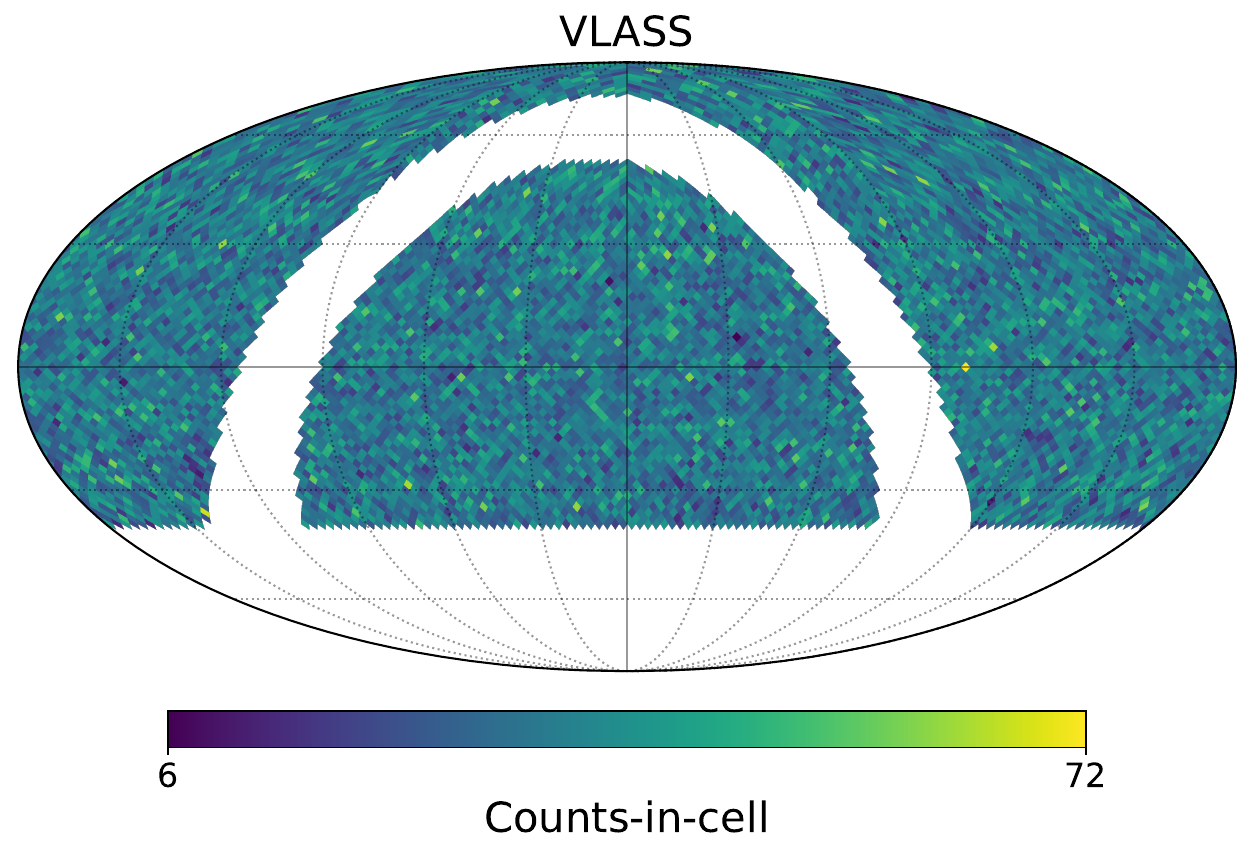}\\
        \includegraphics[width=.32\linewidth]{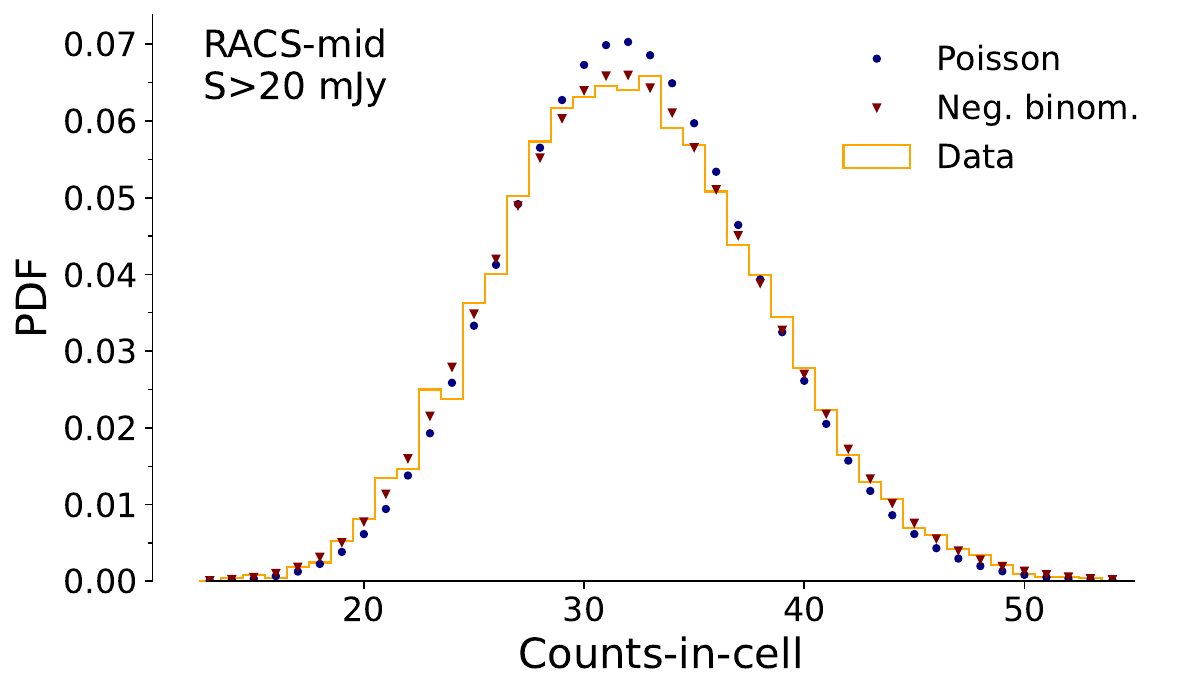}
        \includegraphics[width=.32\linewidth]{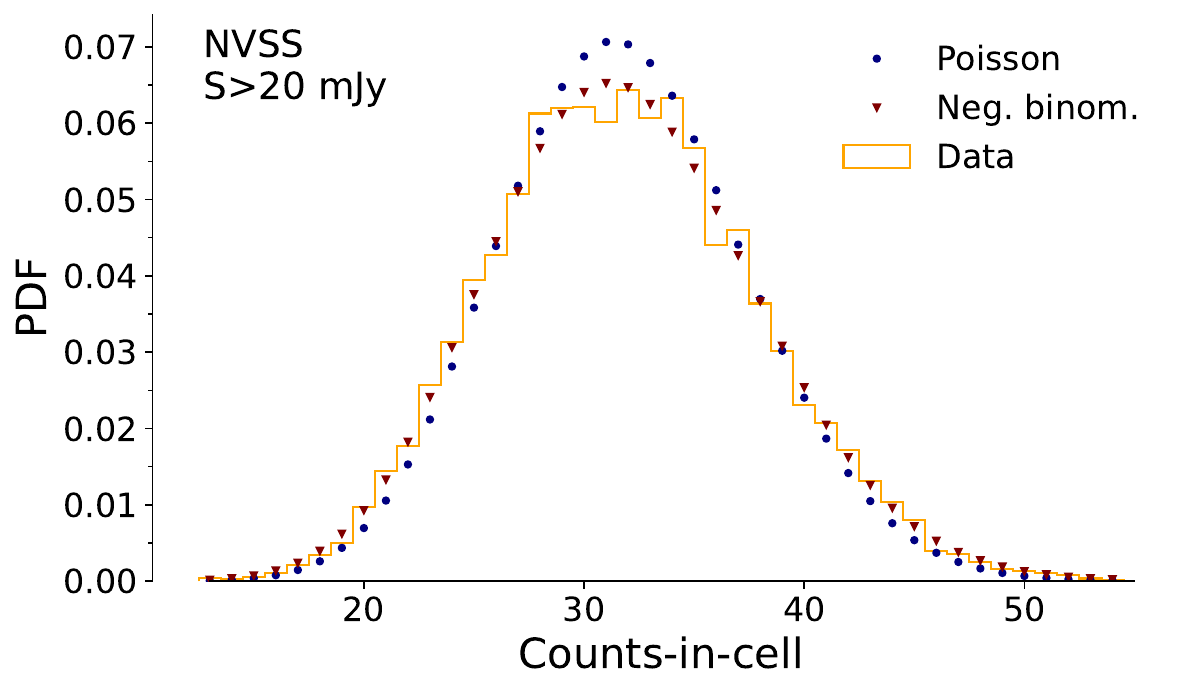}
        \includegraphics[width=.32\linewidth]{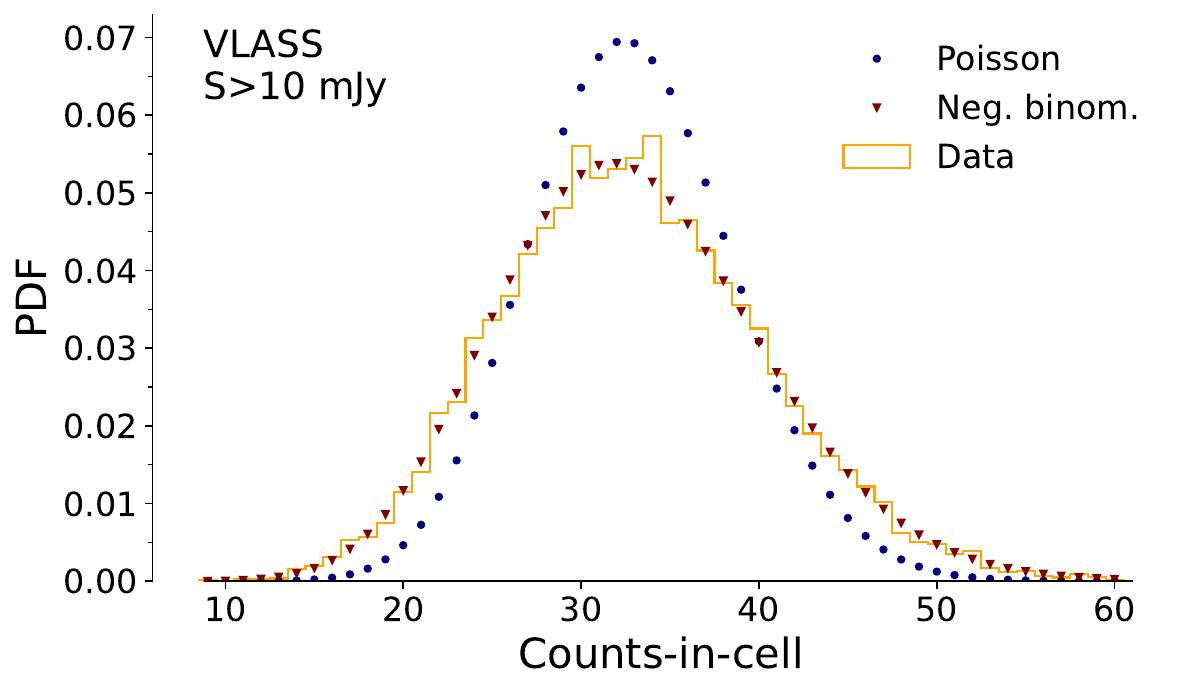}
    \caption{Maps (first and third row) and histograms (second and fourth row) of source counts from six radio continuum surveys and their best-fit Poisson and negative binomial distributions.
\label{fig:data_overview}}
\end{figure*}

\textit{Appendix A: Additional figures \& tables}---This section includes additional figures and tables that provide detailed information on the improvement of the negative binomial distribution for radio surveys, the fitting of the spectral index, and the flux density dependence of the radio dipole direction.
Figure~\ref{fig:data_overview} displays maps and histograms, along with the best-fit Poisson and negative binomial distributions for all six radio surveys. The maps show the applied masks and counts-in-cells for the corresponding flux density cut. The chosen flux density cuts guarantee high degree of completeness and that all six surveys are dominated by radio emission associated with active galactic nuclei. The histogram highlights the enhancement achieved by the negative binomial distribution in comparison to the Poisson distribution in describing the observed counts-in-cells distribution. Fitting parameters are provided in the main text, Table~\ref{tab:comp}.
Table~\ref{tab:alpha} details the cross-matching results for survey pairs. To determine the spectral index of a given radio survey, the chosen mask and flux density cut are applied, and it is matched to another survey without a flux density cut. If multiple matches occur, the flux densities are summed, especially when cross-matching between low and high-resolution surveys. The table also shows the percentage of matches in the overlap of the two surveys. For example, matching LoTSS-DR2 with a 5 mJy flux density cut to NVSS results in 51\% of LoTSS-DR2 sources matching 83\% of NVSS sources in the overlap. The resulting spectral index is $\alpha = -0.73 \pm 0.23$.
Figure~\ref{fig:dec_corners} illustrates ($\alpha_{RA}$, $\delta$) as a function of the flux density cut for all six radio surveys. As discussed in the main text, this helps determine which flux density cuts and surveys are considered robust and which are analyzed with the constrained estimator.
\end{widetext}

\begin{table}[h]
\caption{Percentage of matched radio sources for pairs of surveys and derived mean spectral index, both calculated in their overlap. The third column gives the percentage of matched sources for the first and second surveys, respectively. The given flux density cut is applied to the first survey only, and none applied to the second.
\label{tab:alpha}}
\bgroup\renewcommand{\arraystretch}{1.2}
\begin{tabular}{lccc}\hline\hline
Surveys & $S_\mathrm{min}$ & Percent matched & $\alpha$ \\ \hline
 LoTSS-DR2--NVSS & 5 & 51\% / 83\% & $-0.73\pm0.23$\\
 TGSS--NVSS & 100 & 97\% / 16\% & $-0.76\pm0.18$\\
 RACS-low--NVSS & 20 & 97\% / 25\% & $-0.98\pm0.24$ \\ 
 RACS-low--RACS-mid & 20 & 99\% / 19\% & $-0.95\pm0.23$\\
 RACS-mid--VLASS & 20 & 81\% / 23\% & $-0.81\pm0.30$\\
 FIRST--VLASS\footnote{taken from \citep{gordon2021}, which used the Faint Images of the Radio Sky at Twenty-Centimeters (FIRST; \cite{FIRST}) to calculate $\alpha$.} & - & - & $-0.71$ \\ \hline\hline
\end{tabular}
\egroup
\end{table}

\textit{Appendix B: Systematic effects of radio surveys}--- The systematic effects regarding the large-scale structure of LoTSS-DR2 were discussed and studied in \cite{LoTSS2,hale2023a,pashapour}. Simulations of the LoTSS-DR2 sky coverage and number density show that the dipole retrieval in declination has a small bias but is mainly dominated by a large variance. Simulations using the understood and modeled LoTSS-DR2 systematics, as described by random mocks \cite{hale2023a}, show that a retrieval of the dipole direction is not possible, due to limited sky coverage.
Additionally, we find no cross-correlation between the LoTSS-DR2 map and Galactic synchrotron emission as described by the 408 MHz all-sky Haslam map \cite{haslam1982} from \cite{remazeilles2015}.
\cite{wagenveld2023a} showed that there also is no cross-correlation between the RACS-low number count maps and Galactic synchrotron emission.
For NVSS, \cite{NVSS} provide an extensive discussion of observational systematic effects. 
Systematic effects of NVSS with respect to estimators, survey geometry, flux density cuts, shot noise and more have been studied in \cite{rubart2013,tiwari2016, siewert2021a,secrest2022a}.

\end{document}